\documentclass[conference]{IEEEtran}
\usepackage{algorithmic}
\usepackage{amsmath,amssymb,amsfonts}
\usepackage{tikz}
\usepackage{caption}
\usepackage{cite}
\usepackage{subcaption}
\usepackage{graphicx}
\usetikzlibrary{arrows.meta, positioning}
\usepackage{pgfplots}
\pgfplotsset{compat=newest}
\usetikzlibrary{positioning}
\usepackage{float}
\usepackage{comment}
\pgfplotsset{compat=1.18}
\usepackage{pgfplotstable}
\usepackage{xcolor}
\usetikzlibrary{patterns}
\usepackage{rotating}
\usepackage{balance}
\usepackage{url}

\begin{document}

\title{Antiferromagnetic Tunnel Junctions (AFMTJs) for In-Memory Computing: Modeling and Case Study}

\author{
    \IEEEauthorblockN{Yousuf Choudhary and Tosiron Adegbija\\Department of Electrical and Computer Engineering\\The University of Arizona, USA\\
    Email: ychoudhary@arizona.edu, tosiron@arizona.edu}
}

\maketitle
\vspace{-50pt}
\begin{abstract}
Antiferromagnetic Tunnel Junctions (AFMTJs) enable picosecond switching and femtojoule writes through ultrafast sublattice dynamics. We present the first end-to-end AFMTJ simulation framework integrating multi-sublattice Landau-Lifshitz-Gilbert (LLG) dynamics with circuit-level modeling. SPICE-based simulations show that AFMTJs achieve $\sim$8$\times$ lower write latency and $\sim$9$\times$ lower write energy than conventional MTJs. When integrated into an in-memory computing architecture, AFMTJs deliver 17.5$\times$ average speedup and nearly 20$\times$ energy savings versus a CPU baseline---significantly outperforming MTJ-based IMC. These results establish AFMTJs as a compelling primitive for scalable, low-power computing.
\end{abstract}

\section{Introduction}

Von Neumann architectures create performance bottlenecks from data movement between separate processing and memory units. \textit{In-memory computing (IMC)} addresses this "memory wall" by enabling computation directly within memory arrays, minimizing data movement and its associated costs \cite{li2020memorywall}. However, realizing IMC's potential requires memory technologies that offer high speed, low energy, and high density.

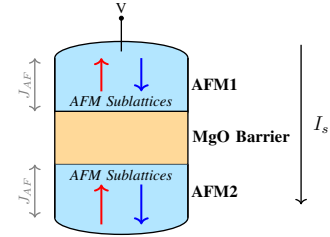
\begin{figure}[!t]
    \vspace{-10pt}
    \centering
    \resizebox{0.5\linewidth}{!}{
    \begin{tikzpicture}[scale=1.4]

        \definecolor{afmblue}{rgb}{0.7,0.9,1}
        \definecolor{barrierorange}{rgb}{1,0.85,0.6}

        \path[fill=afmblue] (0,4) ellipse (1 and 0.25);

        \draw[thick] (0,4) ellipse (1 and 0.25);

        \draw[fill=afmblue, draw=black, thick] (-1,4) -- (-1,3.2) arc[start angle=180,end angle=360,x radius=1,y radius=0.25] -- (1,4);
        \draw[thick] (-1,3.2) -- (-1,4);
        \draw[thick] (1,3.2) -- (1,4);

        \draw[->, very thick, red] (-0.3,3.5) -- (-0.3,4);
        \draw[->, very thick, blue] (0.3,4) -- (0.3,3.5);

        \draw[fill=barrierorange, draw=black, thick] (-1,3.2) rectangle (1,2.4);

        \draw[fill=afmblue, draw=black, thick] (-1,2.4) -- (-1,1.6) arc[start angle=180,end angle=360,x radius=1,y radius=0.25] -- (1,2.4);
        \draw[thick] (-1,1.6) -- (-1,2.4);
        \draw[thick] (1,1.6) -- (1,2.4);

        \draw[->, very thick, red] (-0.3,1.5) -- (-0.3,2.1);
        \draw[->, very thick, blue] (0.3,2.1) -- (0.3,1.5);

        \node at (1.4,3.6) {\textbf{AFM1}};
        \node at (1.8,2.8) {\textbf{MgO Barrier}};
        \node at (1.4,2.0) {\textbf{AFM2}};

        \draw[<->, thick, gray] (-1.3,1.6) -- (-1.3,2.4);
        \node[gray, rotate=90] at (-1.45,2) {\small $J_{AF}$};

        \draw[<->, thick, gray] (-1.3,4) -- (-1.3,3.2);
        \node[gray, rotate=90] at (-1.45,3.6) {\small $J_{AF}$};

        \draw[thick] (0,4.6) -- (0,4.1);
        \draw[fill=black] (0,4.6) circle (0.03);
        \node at (0,4.75) {\small V};

        \draw[->, thick] (2.7,4.2) -- (2.7,1.8); 
        \node at (3,3.0) {\large $I_s$};

        \node at (0,3.35) {\small \textit{AFM Sublattices}};
        \node at (0,2.25) {\small \textit{AFM Sublattices}};

    \end{tikzpicture}
    }
    \caption{AFMTJ device structure. Each AFM layer contains two oppositely aligned sublattices coupled by inter-sublattice exchange ($J_{\text{AF}}$). Applied voltage drives spin-polarized current across the MgO barrier, enabling switching and readout.}
    \label{fig:afmtj_structure}
    \vspace{-10pt}
\end{figure}

Antiferromagnetic Tunnel Junctions (AFMTJs) \cite{shao2024afmtj} are emerging spintronic devices that provide picosecond-scale switching \cite{cheng2015ultrafast}, low write energy \cite{deng2017qmtj}, high packing density for near-terahertz IMC architectures, and robustness against external magnetic fields due to near-zero net magnetization. Unlike conventional Magnetic Tunnel Junctions (MTJs), which switch via net magnetization toggling, AFMTJs switch through reorientation of antiparallel magnetic sublattices (Fig. \ref{fig:afmtj_structure}), enabling sub-100 ps switching far exceeding the nanosecond-scale dynamics of MTJs (Table \ref{tab:mtj_vs_afmtj}).

Despite their promise, AFMTJs remain unexplored in computing architectures. Prior work has focused on physical characterization \cite{deng2017qmtj,ikeda2010perpendicular,shao2024afmtj,ranjbar2015heusler,usami2021damping} without demonstrating dynamic models integrated into circuit-level simulations or functional architectures.

This paper makes three main contributions: \textbf{(1)} We develop the first compact AFMTJ model for in-memory computing, capturing dual-sublattice Landau-Lifshitz-Gilbert dynamics and inter-sublattice exchange coupling; \textbf{(2)} We validate our model against experimental tunneling magnetoresistance (TMR) and switching characteristics; \textbf{(3)} We demonstrate system-level IMC gains of 17.5$\times$ average speedup and $\sim$20$\times$ energy savings over a CPU baseline, significantly outperforming MTJ-based alternatives. This work lays a foundation for future architectural exploration of AFMTJ-based systems for ultrafast, energy-efficient, and scalable in-memory computing.

\begin{table}[t!]
\centering
\caption{Comparison of MTJ and AFMTJ characteristics.}
\renewcommand{\arraystretch}{1.15}
\setlength{\tabcolsep}{4pt}
\scriptsize
\begin{tabular}{|p{2.4cm}|p{2.6cm}|p{2.6cm}|}
\hline
\textbf{Property} & \textbf{MTJ} & \textbf{AFMTJ} \\
\hline
Sublattice Structure & Single ferromagnetic (FM) layer & Two antiparallel AFM sublattices\\
\hline
TMR Ratio & 80--120\% \cite{lee2015senseamp, zhou2023pma} & Up to 500\% \cite{shao2024afmtj} \\
\hline
Switching Time & 1--2 ns \cite{lee2015senseamp} & 10--100 ps \cite{shao2024afmtj, usami2021damping} \\
\hline
Write Energy & 300 fJ \cite{patel2016field} & 20--100 fJ [this work] \\
\hline
Field Sensitivity & High & Low (inherent robustness due to zero net magnetization ) \\
\hline
\end{tabular}
\label{tab:mtj_vs_afmtj}
\end{table}

\section{AFMTJ Device Model} \label{sec:model}
We extend the UMN MTJ SPICE framework \cite{kim2015mtj} to capture the dual-sublattice AFMTJ physics\footnote{Our AFMTJ SPICE model is available at:\\ \url{https://github.com/yousufc890/AFMTJ_Model}}. Each cell instantiates two dynamically coupled magnetization vectors, 
$\mathbf{M}_1$ and $\mathbf{M}_2$, evolving under modified Landau-Lifshitz-Gilbert \cite{lakshmanan2011fascinating_llg} equations:

\[
\frac{d\mathbf{M}_i}{dt} = -\gamma \mathbf{M}_i \times \mathbf{H}_{\text{eff},i} + \alpha \mathbf{M}_i \times \frac{d\mathbf{M}_i}{dt} + \boldsymbol{\tau}_{\text{STT},i} + \boldsymbol{\tau}_{\text{ex},i}
\]

where $\mathbf{H}_{\text{eff},i}$ includes the demagnetizing, anisotropy, thermal, and inter-sublattice exchange fields. The exchange torque terms $\boldsymbol{\tau}_{\text{ex},1} = -J_{\mathrm{AF}} \, \mathbf{M}_1 \times \mathbf{M}_2$ couple the sublattices, enabling mutual dynamic feedback. An adaptive fourth-order Runge-Kutta integrator (0.1 ps base step) evolves all six magnetization components. 

\subsection{Validation and Comparison}

We validated our model against fabricated AFMTJs\cite{Chou2024LargeSpin, gurung2024spinpolarization, chen2024twist}, achieving comparable TMR values ($\sim$80\%) and picosecond switching dynamics. Table \ref{tab:mtj_vs_afmtj} summarizes key differences from conventional MTJs and Table~\ref{tab:afmtj_params} summarizes the key material parameters used in this work. 

\begin{table}[t!]
\centering
\caption{AFMTJ Parameters Used}
\vspace{-5pt}
\scriptsize
\begin{tabular}{|l|c|l|}
\hline
\textbf{Parameter} & \textbf{Value} & \textbf{Description} \\
\hline
P0 & 0.8 & Polarization Factor \\
$\alpha$ & 0.01 & Damping Factor \\
Ms0 & $600~\mathrm{emu/cm^3}$ & Saturation Magnetization \\
$J_{\mathrm{AF}}$ & 5e-3 & Inter-sublattice Exchange Coupling Constant \\
lx, ly & 45nm & Lateral dimensions of free layer \\
lz & 0.45nm & Thickness of free layer \\
\hline
\end{tabular}
\label{tab:afmtj_params} 
\vspace{0.6em}
\begin{minipage}{\columnwidth}
\scriptsize
Values were selected based on known properties of AFM materials. AFMTJ cell dimensions were kept consistent with the original UMN MTJ model \cite{kim2015mtj,ranjbar2015heusler,usami2021damping,gurung2024spinpolarization,prapplied2019,jungwirth2016afmspintronics,gibertini2019magnetic2d}.
\end{minipage}
\vspace{-5pt}
\end{table}

\section{AFMTJ-Based In-Memory Computing Architecture}

\subsection{System Integration}
\begin{figure}[t!]
\centering
\resizebox{0.9\columnwidth}{!}{
\begin{tikzpicture}[node distance=1.6cm and 1.3cm, every node/.style={font=\large}]
  \node[draw, fill=green!20, minimum width=2cm, minimum height=1.2cm] (cpu) {CPU};

  \node[draw, fill=green!15, right=of cpu, minimum width=3.3cm, minimum height=2.4cm] (l1) {};
  \node at (l1.north) [yshift=-0.25cm] {\textbf{L1 Cache}};
  \node[draw, fill=gray!30, minimum width=1cm, minimum height=0.8cm] at ([xshift=-0.9cm,yshift=0.3cm]l1.center) (c1) {C1};
  \node[draw, fill=gray!30, minimum width=1cm, minimum height=0.8cm] at ([xshift=0.9cm,yshift=0.3cm]l1.center) (c2) {C2};
  \node at ([yshift=-0.9cm]l1.center) {AFMTJ Logic};

  \node[draw, fill=green!15, right=of l1, minimum width=3.3cm, minimum height=2.4cm] (l2) {};
  \node at (l2.north) [yshift=-0.25cm] {\textbf{L2 Cache}};
  \node[draw, fill=gray!30, minimum width=1cm, minimum height=0.8cm] at ([xshift=-0.9cm,yshift=0.3cm]l2.center) (c3) {C3};
  \node[draw, fill=gray!30, minimum width=1cm, minimum height=0.8cm] at ([xshift=0.9cm,yshift=0.3cm]l2.center) (c4) {C4};
  \node at ([yshift=-0.9cm]l2.center) {AFMTJ Logic};

  \node[draw, fill=green!15, right=of l2, minimum width=3.3cm, minimum height=2.4cm] (mem) {};
  \node at (mem.north) [yshift=-0.25cm] {\textbf{Main Memory}};
  \node[draw, fill=gray!30, minimum width=1cm, minimum height=0.8cm] at ([xshift=-0.9cm,yshift=0.3cm]mem.center) (c5) {C5};
  \node[draw, fill=gray!30, minimum width=1cm, minimum height=0.8cm] at ([xshift=0.9cm,yshift=0.3cm]mem.center) (c6) {C6};
  \node at ([yshift=-0.9cm]mem.center) {AFMTJ Logic};

  \draw[<->, thick] (cpu) -- (l1);
  \draw[<->, thick] (l1) -- (l2);
  \draw[<->, thick] (l2) -- (mem);

  \foreach \i in {c1, c2, c3, c4, c5, c6} {
    \draw[<->, thick] (\i.south) -- ++(0,-0.5);
  }

  \draw[thick] ([yshift=-0.5cm]c1.south) -- ([yshift=-0.5cm]c6.south);

  \node at ([yshift=-0.5cm]l1.south) {\textit{PiC}};
  \node at ([yshift=-0.5cm]mem.south) {\textit{PiM}};
\end{tikzpicture}
}
\vspace{-7pt}
\caption{High-level system architecture showing hierarchical AFMTJ-based in-memory compute. AFMTJ subarrays (C1--C6) serve as both memory (data) and compute (logic) blocks within L1, L2, and main memory, enabling processing in cache (PiC) and processing in memory (PiM).}
\label{fig:afmtj_system}
\vspace{-15pt}
\end{figure}
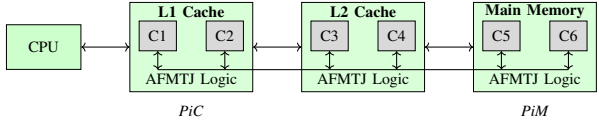

To explore AFMTJ's system-level benefits, we integrate it into a hierarchical IMC architecture \cite{gajaria2024chime} to replace MTJs. Fig. \ref{fig:afmtj_system} illustrates the hierarchical IMC architecture, wherein AFMTJ subarrays embedded at L1, L2, and main memory serve as both non-volatile storage and fine-grained logic operators (XOR, NAND). Multi-row activation and charge sharing implement bit-line computing, with sense amplifiers resolving logic outcomes from magnetization-dependent current differentials.

\subsection{Cell Operations}
The AFMTJ cells support three modes:
\textbf{\textit{write}} (a spin-polarized current alters magnetization); \textbf{\textit{read}} (the TMR is sensed through the tunnel barrier to resolve the stored state); and \textbf{\textit{logic}} (conditional voltage pulses on multiple rows drive in-place Boolean operations such as NAND or XOR). A lightweight controller orchestrates the operations, exploiting AFMTJ's picosecond switching for pipelined execution.

\section{Evaluation}

\subsection{Experimental Setup}
We used Synopsys HSPICE for device simulations. To explore AFMTJ as a drop-in replacement for MTJ in a system-level evaluation, we used the calibrated AFMTJ and a baseline MTJ \cite{kim2015mtj} in an identical IMC hierarchy. The baseline CPU is a 2 GHz ARM Cortex-A72 with 32 KB L1, 1 MB L2, and 8 GB main memory. Workloads include binarized neural network (\textit{bnn}), image grayscale (\textit{img-grayscale}), image thresholding (\textit{img-threshold}), multiply accumulate (\textit{mac}), matrix addition (\textit{mat\_add}), and root mean square error (\textit{rmse}). 

\subsection{Device-Level Results} \label{sec:device-results}
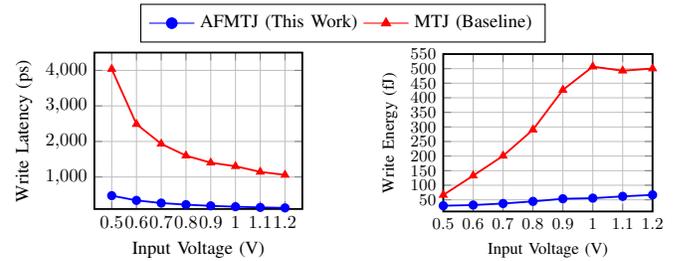
\begin{figure}[t!]
\centering
\vspace{-5pt}

\begin{tikzpicture}
\begin{axis}[
    hide axis,
    xmin=0, xmax=1,
    ymin=0, ymax=1,
    width=0.6\columnwidth,
    legend style={at={(0.5,1.1)}, anchor=south, legend columns=2, font=\scriptsize},
    legend cell align={left}
]
\addlegendimage{blue, mark=*}
\addlegendentry{AFMTJ (This Work)}
\addlegendimage{red, mark=triangle*}
\addlegendentry{MTJ (Baseline)}
\end{axis}
\end{tikzpicture}

\begin{subfigure}[t]{0.45\columnwidth}
  \centering
  \resizebox{\linewidth}{!}{\centering

\begin{tikzpicture}
  \begin{axis}[
    scale only axis,
    xlabel={Input Voltage (V)},
    ylabel={Write Latency (ps)},
    grid=major,
    width=\linewidth,
    height=3cm,
    mark size=2pt,
    line width=1pt,
    xtick={0.5,0.6,0.7,0.8,0.9,1.0,1.1,1.2},
    yticklabel style={/pgf/number format/.cd, fixed, precision=0},
    ymin=100, ymax=4500,
  ]
    \addplot[
      color=blue,
      mark=*,
    ]
    coordinates {
      (0.5, 475.2)
      (0.6, 341.4)
      (0.7, 267.2)
      (0.8, 222.0)
      (0.9, 188.1)
      (1.0, 163.6)
      (1.1, 144.7)
      (1.2, 130.7)
    };

    \addplot[
      color=red,
      mark=triangle*,
    ]
    coordinates {
      (0.5, 4037)
      (0.6, 2487)
      (0.7, 1933)
      (0.8, 1601)
      (0.9, 1402)
      (1.0, 1302)
      (1.1, 1144)
      (1.2, 1059)
    };
  \end{axis}
\end{tikzpicture}

\label{fig:write_vs_vmtj}}
  \caption{Write Latency}
  \label{fig:write_latency}
\end{subfigure}
\hfill
\begin{subfigure}[t]{0.45\columnwidth}
  \centering
  \resizebox{\linewidth}{!}{\centering

\begin{tikzpicture}
\begin{axis}[
    scale only axis,
    width=\linewidth,
    height=3cm,
    xlabel={Input Voltage (V)},
    ylabel={Write Energy (fJ)},
    xmin=0.5, xmax=1.2,
    ymin=10, ymax=550,
    xtick={0.5,0.6,0.7,0.8,0.9,1.0,1.1,1.2},
    ytick={50,100,...,550},
    grid=both,
    grid style={line width=.1pt, draw=gray!30},
    major grid style={line width=.2pt,draw=gray!50},
    tick label style={font=\small},
    label style={font=\small},
    line width=1pt,
    mark size=2pt,
]

\addplot[
    color=blue,
    mark=*,
]
coordinates {
    (0.5, 29.70)
    (0.6, 32.02)
    (0.7, 37.26)
    (0.8, 44.59)
    (0.9, 53.56)
    (1.0, 55.65)
    (1.1, 61.78)
    (1.2, 66.93)
};

\addplot[
    color=red,
    mark=triangle*,
]
coordinates {
  (0.5, 66.58)
  (0.6, 132.97)
  (0.7, 200.85)
  (0.8, 290.43)
  (0.9, 426.88)
  (1.0, 506.94)
  (1.1, 492.84)
  (1.2, 500.29)
};
\end{axis}
\end{tikzpicture}

\label{fig:energy_vs_vmtj}}
  \caption{Write Energy}
  \label{fig:write_energy}
\end{subfigure}

\vspace{-4pt}
\caption{Write (a) latency and (b) energy comparison of AFMTJ vs. MTJ across input voltages.}
\label{fig:combined_latency_energy}
 \vspace{-5pt}
\end{figure}

Fig. \ref{fig:combined_latency_energy} shows performance across input voltages. At 1.0 V, AFMTJ achieves 164 ps write latency (vs. $\sim$1400 ps for MTJ) and 55.7 fJ write energy (vs. $\sim$480 fJ), representing $\sim$8$\times$ and $\sim$9$\times$ improvements, respectively. Switching latency drops from 65 ps at 0.5 V to 20 ps at 1.2 V.

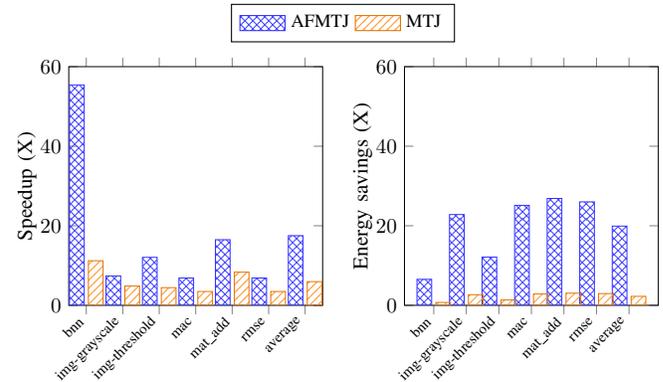
\begin{figure}[t!]
\centering

\begin{tikzpicture}
\begin{axis}[
    hide axis,
    xmin=0, xmax=1,
    ymin=0, ymax=1,
    width=\columnwidth,
    legend style={at={(0.5,1.05)}, anchor=south, legend columns=2, font=\scriptsize},
    legend cell align={left}
]

\addlegendimage{ybar, area legend, bar width=6pt, pattern=crosshatch, pattern color=blue!70, draw=blue!80}
\addlegendentry{AFMTJ}

\addlegendimage{ybar, area legend, bar width=6pt,
pattern=north east lines, pattern color=orange!80, draw=orange!90!black}
\addlegendentry{MTJ}

\end{axis}
\end{tikzpicture}

\vspace{4pt}

\begin{subfigure}[t]{0.48\columnwidth}
  \centering
  \resizebox{\linewidth}{!}{\centering
\begin{tikzpicture}
\begin{axis}[
scale only axis,
  ybar,
  bar width=7pt,
  width=\linewidth,
  height=4cm,
  ymin=0, ymax=60,                
  enlarge x limits=0.08,
  symbolic x coords={bnn,img-grayscale,img-threshold,mac,mat\_add,rmse,average},
  xtick=data,
  xticklabel style={rotate=45, anchor=east, font=\scriptsize},
  ylabel={Speedup (X)},
  legend style={at={(0.5,-0.35)}, anchor=north, legend columns=-1, font=\scriptsize},
  every axis y label/.append style={yshift=-5pt}
]

\addplot+[ybar, bar width=7pt, pattern=crosshatch, pattern color=blue!70, draw=blue!80] coordinates {
  (bnn,55.42) (img-grayscale,7.40) (img-threshold,12.06)
  (mac,6.87) (mat\_add,16.50) (rmse,6.87) (average,17.52)
};

\addplot+[ybar, bar width=7pt, pattern=north east lines, pattern color=orange!80, draw=orange!90!black] coordinates {
  (bnn,11.21) (img-grayscale,4.85) (img-threshold,4.44)
  (mac,3.48) (mat\_add,8.34) (rmse,3.48) (average,5.96)
};

\end{axis}
\end{tikzpicture}
}
  \vspace{-20pt}
  \caption{Latency}
  \label{fig:workload_latency}
\end{subfigure}%
\hspace{0.01\columnwidth}
\begin{subfigure}[t]{0.48\columnwidth}
  \centering
  \resizebox{\linewidth}{!}{\centering
\begin{tikzpicture}
\begin{axis}[
  scale only axis,
  ybar,
  bar width=7pt,
  width=\linewidth,
  height=4cm,
  ymin=0, ymax=60,                
  enlarge x limits=0.15,
  symbolic x coords={bnn,img-grayscale,img-threshold,mac,mat\_add,rmse,average},
  xtick=data,
  xticklabel style={rotate=45, anchor=east, font=\scriptsize},
  ylabel={Energy savings (X)},
  legend style={at={(0.5,-0.6)}, anchor=north, legend columns=-1, font=\scriptsize},
  every axis y label/.append style={yshift=-5pt}
]

\addplot+[ybar, bar width=7pt, pattern=crosshatch, pattern color=blue!70, draw=blue!80] coordinates {
  (bnn,6.54) (img-grayscale,22.82) (img-threshold,12.13)
  (mac,25.14) (mat\_add,26.87) (rmse,26.01) (average,19.92)
};

\addplot+[ybar, bar width=7pt, pattern=north east lines, pattern color=orange!80, draw=orange!90!black] coordinates {
  (bnn,0.72) (img-grayscale,2.67) (img-threshold,1.35)
  (mac,2.85) (mat\_add,3.05) (rmse,2.95) (average,2.26)
};

\end{axis}
\end{tikzpicture}
}
  \vspace{-20pt}
  \caption{Energy}
  \label{fig:workload_energy}
\end{subfigure}
\vspace{-4pt}
\caption{System-level (a) latency speedup and (b) energy savings of AFMTJ- and MTJ-based hierarchical IMC architecture versus the CPU baseline across workloads.}
\label{fig:workload_combined_vertical}
\vspace{-15pt}
\end{figure}

\subsection{System-Level Results} \label{sec:imc_evaluation}


Fig. \ref{fig:workload_combined_vertical} summarizes the performance of AFMTJ- and MTJ-based IMC architectures against the CPU baseline. AFMTJ-based IMC achieves \textbf{17.5$\times$ average speedup} and \textbf{19.9$\times$ average energy savings}, compared to 6$\times$ and 2.3$\times$ for MTJ-based IMC. Write-intensive workloads (\textit{bnn}: 55.4$\times$; \textit{mat\_add}: 16.5$\times$) show the largest gains. 

\section{Conclusion}

We presented the first SPICE-based AFMTJ simulation framework capturing dual-sublattice dynamics and demonstrated its integration into in-memory computing. AFMTJs achieve $\sim$8$\times$ lower write latency and $\sim$9$\times$ lower energy than MTJs at the device level, translating to 17.5$\times$ speedup and $\sim$20$\times$ energy savings at the system level. Future work includes compact Verilog-A models, fabrication validation, and architecture-level studies targeting edge AI, real-time signal processing, and other high-impact applications. 

\section*{Acknowledgment}
This work was partially supported by NSF Grant 2425567.

\bibliographystyle{IEEEtran}
\bibliography{references}

@article{lakshmanan2011fascinating_llg,
  title={The fascinating world of the Landau--Lifshitz--Gilbert equation: an overview},
  author={Lakshmanan, Muthusamy},
  journal={Philosophical Transactions of the Royal Society A: Mathematical, Physical and Engineering Sciences},
  volume={369},
  number={1939},
  pages={1280--1300},
  year={2011},
  publisher={The Royal Society Publishing}
}

@inproceedings{kim2015mtj,
  author    = {J. Kim and A. Chen and B. Behin-Aein and S. Kumar and J. P. Wang and C. H. Kim},
  title     = {A Technology-Agnostic MTJ SPICE Model with User-Defined Dimensions for STT-MRAM Scalability Studies},
  booktitle = {Custom Integrated Circuits Conference (CICC)},
  year      = {2015},
  month     = sep
}

@article{shao2024afmtj,
  author  = {D. F. Shao and E. Y. Tsymbal},
  title   = {Antiferromagnetic tunnel junctions for spintronics},
  journal = {npj Spintronics},
  volume  = {2},
  pages   = {13},
  year    = {2024},
  doi     = {10.1038/s44306-024-00014-7}
}

@article{ranjbar2015heusler,
  author  = {R. Ranjbar and K. Suzuki and A. Sugihara and T. Miyazaki and Y. Ando and S. Mizukami},
  title   = {Engineered Heusler Ferrimagnets with a Large Perpendicular Magnetic Anisotropy},
  journal = {Materials},
  volume  = {8},
  number  = {9},
  pages   = {6531--6542},
  year    = {2015},
  doi     = {10.3390/ma8095320}
}

@article{usami2021damping,
  author  = {T. Usami and M. Itoh and T. Taniyama},
  title   = {Temperature dependence of the effective Gilbert damping constant of FeRh thin films},
  journal = {AIP Advances},
  volume  = {11},
  number  = {4},
  pages   = {045302},
  year    = {2021},
  doi     = {10.1063/5.0039577}
}

@article{gibertini2019magnetic2d,
  author  = {M. Gibertini and M. Koperski and A. F. Morpurgo and others},
  title   = {Magnetic 2D materials and heterostructures},
  journal = {Nat. Nanotechnol.},
  volume  = {14},
  pages   = {408--419},
  year    = {2019},
  doi     = {10.1038/s41565-019-0438-6}
}

@article{jungwirth2016afmspintronics,
  author  = {T. Jungwirth and X. Marti and P. Wadley and others},
  title   = {Antiferromagnetic spintronics},
  journal = {Nature Nanotech},
  volume  = {11},
  pages   = {231--241},
  year    = {2016},
  doi     = {10.1038/nnano.2016.18}
}

@article{cheng2015ultrafast,
  author  = {R. Cheng and M. W. Daniels and J. Zhu and D. Xiao},
  title   = {Ultrafast Switching of Antiferromagnets via Spin-transfer Torque},
  journal = {Phys. Rev. B},
  volume  = {91},
  pages   = {064423},
  year    = {2015},
  doi     = {10.1103/PhysRevB.91.064423}
}

@article{deng2017qmtj,
  author  = {J. Deng and G. Liang and G. Gupta},
  title   = {Ultrafast and low-energy switching in voltage-controlled elliptical pMTJ},
  journal = {Scientific Reports},
  volume  = {7},
  pages   = {16562},
  year    = {2017},
  doi     = {10.1038/s41598-017-16292-7}
}

@article{prapplied2019,
  author = {Zhanran Wang and Bo Bian and Lei Zhang and Zhizhou Yu},
  title   = {Mn3Sn-based noncollinear antiferromagnetic tunnel junctions with bilayer boron nitride tunnel barriers},
  journal = {Applied Physics Letters},
  year    = {2019},
  doi     = {10.1063/5.0234130}
}

@article{li2020memorywall,
  author  = {S. Li and et al.},
  title   = {Architecting to break the memory wall for emerging workloads},
  journal = {IEEE Micro},
  volume  = {40},
  number  = {4},
  pages   = {68--76},
  year    = {2020}
}

@inproceedings{gajaria2024chime,
  author    = {D. Gajaria and T. Adegbija and K. Gomez},
  title     = {CHIME: Energy-Efficient STT-RAM-Based Concurrent Hierarchical In-Memory Processing},
  booktitle = {IEEE 35th International Conference on Application-specific Systems, Architectures and Processors (ASAP)},
  year      = {2024},
  pages     = {228--236},
  doi       = {10.1109/ASAP61560.2024.00053}
}

@article{lee2015senseamp,
  author  = {H. Lee and et al.},
  title   = {Design of a Fast and Low-Power Sense Amplifier and Writing Circuit for High-Speed MRAM},
  journal = {IEEE Transactions on Magnetics},
  volume  = {51},
  number  = {5},
  pages   = {1--7},
  year    = {2015},
  doi     = {10.1109/TMAG.2014.2367130}
}

@article{gurung2024spinpolarization,
  author  = {G. Gurung and M. Elekhtiar and Q. Q. Luo and et al.},
  title   = {Nearly perfect spin polarization of noncollinear antiferromagnets},
  journal = {Nature Communications},
  volume  = {15},
  pages   = {10242},
  year    = {2024},
  doi     = {10.1038/s41467-024-54526-1}
}

@article{zhou2023pma,
  author       = {Zhou, B. and Khanal, P. and Benally, O. J. and Lin, J. and Zhang, J. and Liu, Y. and Jiang, Y. and Wang, C. and Li, X. and Xiao, G.},
  title        = {Perpendicular magnetic anisotropy, tunneling magnetoresistance and spin-transfer torque effect in magnetic tunnel junctions with Nb layers},
  journal      = {Scientific Reports},
  volume       = {13},
  pages        = {3454},
  year         = {2023},
  publisher    = {Nature Publishing Group},
  doi          = {10.1038/s41598-023-29752-0},
  url          = {https://doi.org/10.1038/s41598-023-29752-0}
}

@article{patel2016field,
  author = {Patel, Ravi and Guo, Xiaochen and Guo, Qing and Ipek, Engin and Friedman, Eby G.},
  title = {Reducing Switching Latency and Energy in STT‑MRAM Caches with Field‑Assisted Writing},
  journal = {IEEE Transactions on Very Large Scale Integration (VLSI) Systems},
  volume = {24},
  number = {1},
  pages = {129--138},
  year = {2016},
  doi = {10.1109/TVLSI.2015.2401577}
}

@article{ikeda2010perpendicular,
  title={A perpendicular-anisotropy CoFeB–MgO magnetic tunnel junction},
  author={Ikeda, S. and Miura, K. and Yamamoto, H. and Mizunuma, K. and Gan, H. D. and Endo, M. and Kanai, S. and Hayakawa, J. and Matsukura, F. and Ohno, H.},
  journal={Nature Materials},
  volume={9},
  number={9},
  pages={721--724},
  year={2010},
  publisher={Nature Publishing Group},
  doi={10.1038/nmat2804}
}

@article{chen2024twist,
  title={Twist-assisted all‑antiferromagnetic tunnel junction in the atomic limit},
  author={Chen, Yuliang and Samanta, Kartik and Shahed, Naafis A. and Zhang, Haojie and Fang, Chi and Ernst, Arthur and Tsymbal, Evgeny Y. and Parkin, Stuart S. P.},
  journal={Nature},
  volume={632},
  pages={1045–1051},
  year={2024},
  doi={10.1038/s41586-024-07818-x}
}

@article{Chou2024LargeSpin,
  author    = {Chou, T. and Ghosh, S. and McGoldrick, B. C. and Nguyen, T. and Gurung, G. and Tsymbal, E. Y. and Li, M. and Mkhoyan, K. A. and Liu, L.},
  title     = {Large Spin Polarization from Symmetry-Breaking Antiferromagnets in Antiferromagnetic Tunnel Junctions},
  journal   = {Nature Communications},
  year      = {2024},
  volume    = {15},
  pages     = {7840},
  doi       = {10.1038/s41467-024-52208-6},
  url       = {https://doi.org/10.1038/s41467-024-52208-6}
}
\balance

\end{document}